\documentclass[10pt,superscriptaddress,twocolumn,showpacs,prb,aps]{revtex4-1}

\usepackage{graphicx}
\usepackage{dcolumn}
\usepackage{bm}
\usepackage{xspace}
\usepackage{multirow}
\usepackage{natbib}
\usepackage{color}
\usepackage{makecell}
\usepackage{ulem}
\usepackage{soul}

\newcommand{\RUC}{Department of Physics, Renmin University of China, Beijing 100872, China}
\newcommand{\BJLab}{Beijing Key Laboratory of Opto-electronic Functional Materials $\textsl{\&}$ Micro-nano Devices, Renmin University of China, Beijing, China}
\newcommand{\IOP}{Beijing National Laboratory for Condensed Matter Physics, and Institute of Physics, Chinese Academy of Sciences, Beijing 100190, China}
\newcommand{\CenterQM}{Collaborative Innovation Center of Quantum Matter, Beijing, China}

\newcommand{\ZST}{ZrSnTe}

\begin{document}

\title{Emergence of topological bands on the surface of {\ZST} crystal}

  \author{R. Lou}
    \thanks{These authors contributed equally to this work.}
    \affiliation{\RUC}
    \affiliation{\BJLab}
  \author{J.-Z. Ma}
    \thanks{These authors contributed equally to this work.}
  \author{Q.-N. Xu}
    \thanks{These authors contributed equally to this work.}
  \author{B.-B. Fu}
  \author{L.-Y. Kong}
  \author{Y.-G. Shi}
    \affiliation{\IOP}
  \author{P. Richard}
    \affiliation{\IOP}
    \affiliation{\CenterQM}
  \author{H.-M. Weng}
    \email{hmweng@iphy.ac.cn}
    \affiliation{\IOP}
    \affiliation{\CenterQM}
  \author{Z. Fang}
    \affiliation{\IOP}
    \affiliation{\CenterQM}
  \author{S.-S. Sun}
  \author{Q. Wang}
  \author{H.-C. Lei}
    \email{hlei@ruc.edu.cn}
    \affiliation{\RUC}
    \affiliation{\BJLab}
  \author{T. Qian}
    \email{tqian@iphy.ac.cn}
    \affiliation{\IOP}
  \author{H. Ding}
    \affiliation{\IOP}
    \affiliation{\CenterQM}
  \author{S.-C. Wang}
    \email{scw@ruc.edu.cn}
    \affiliation{\RUC}
    \affiliation{\BJLab}

\begin{abstract}
  By using angle-resolved photoemission spectroscopy combined with first-principles calculations, we reveal that the topmost unit
  cell of {\ZST} crystal hosts two-dimensional (2D) electronic bands of topological insulator (TI) state, though such a TI state
  is defined with a curved Fermi level instead of a global band gap. Furthermore, we find that by modifying the dangling bonds on
  the surface through hydrogenation, this 2D band structure can be manipulated so that the expected global energy gap is most likely
  to be realized. This facilitates the practical applications of 2D TI in heterostructural devices and those with surface decoration
  and coverage. Since {\ZST} belongs to a large family of compounds having the similar crystal and band structures, our findings shed
  light on identifying more 2D TI candidates and superconductor-TI heterojunctions supporting topological superconductors.
\end{abstract}


{\maketitle}

After nearly a decade of intensive studies, the field of topological insulators (TIs) has led to remarkable achievements
\cite{Hasan2010,Qi2011,Weng2014}, such as the discoveries of two-dimensional (2D) \cite{Bernevig2006,Konig2007,Knez2011}
and three-dimensional (3D) TIs \cite{Chen2009}, quantum anomalous Hall effect (or Chern insulators) \cite{Yu2010,Chang2013,
Wengadv2015}, topological crystalline insulators \cite{Fu2011,Heish2012NC} and even topological semimetals like Dirac
semimetals \cite{Wang2012&2013,Liu2014,Liu2014a} and Weyl semimetals \cite{Weng2015Weyl,Lv2015,Huang2015,Xu2015Weyl,Lv2015a,
Xu2015Weyla}. However, when looking back one finds that the study of 2D TIs is still at the early stage it was when it
ignited the whole field. 2D TIs have more promising potential applications than their 3D cousins \cite{Ando2013}. The lack
of experimentally suitable 2D TI materials is the main problem. A favorable 2D TI is expected to have a quite large band
gap to be operable under easily accessible temperature and to be prepared easily \cite{Weng2014}. Considerable theoretical
efforts have predicted many 2D TI candidates in recent years \cite{Weng2014PRX,BiRhI,Bi4Br4,KHgSb,Tin}, but few of them
has been confirmed experimentally.

Very recently, Xu $et$ $al$. have proposed that ZrSiO monolayer is a 2D TI with a band gap up to 30 meV, and that its
isostructural compounds $WHM$ ($W$ = Zr, Hf, or La, $H$ = Si, Ge, Sn, or Sb, and $M$ = O, S, Se, and Te) possess similar
electronic structures~\cite{Xu2015}. It is also proposed that if the inherent spin-orbit coupling (SOC) is neglected,
they are node-line semimetals \cite{Xu2015,ZrSiS2015}. In this work, we report systematic angle-resolved photoemission
spectroscopy (ARPES) measurements on {\ZST} single crystals. By comparing with first-principles calculations, we reveal
that the topmost unit cell on the (001) surface hosts a 2D electronic structure, which is significantly consistent with
that of the proposed 2D TI {\ZST} monolayer. However, such a 2D TI state is defined with a curved Fermi level ($E_F$)
instead of a global band gap to ensure a finite gap at each crystal momentum $k$ point. We further prove that such
topological bands can be engineered by modifying the dangling bonds on the terminating layer. The hydrogenation of the
dangling bonds of Zr 4$d$ orbitals tends to align the band gaps to the same energy level. This observation is excellently
reproduced by our slab calculations, which predict that a global band gap can be reached at 0.5 eV above the chemical
potential as all the Zr ions on the terminating layer are bonded with hydrogen. Our findings suggest that the $WHM$ series
is a plausible platform to support 2D TI with nontrivial electronic bands, for which the surface decoration, or interface
of heterostructure can realize a global band gap, which is more suitable for the practical device application than
free-standing 2D systems.

High-quality single crystals of {\ZST} were grown by the Te flux method. ARPES measurements were performed at the ``Dreamline''
beamline of the Shanghai Synchrotron Radiation Facility (SSRF) with a Scienta D80 analyzer, and at the beamline 13U of the National
Synchrotron Radiation Laboratory (NSRL) with a Scienta R4000 analyzer. The samples were cleaved $in$ $situ$ and measured at $T$ =
20--40 K in a vacuum better than 5$\times$10$^{-11}$ Torr. The energy and angular resolutions were set to 15 meV and 0.2$^{\circ}$,
respectively. The ARPES data were collected using horizontally polarized light with a vertical analyzer slit. The hydrogen adsorption
process was controlled by exposing samples in vacuum for a longer time at the lowest temperature 20 K. In contrast, no noticeable
changes in the spectra were observed during the regular measurements at $T$ = 40 K. Most of the first-principles calculations were
carried out using the Vienna $ab$ $initio$ simulation package (VASP) \cite{Kresse1996a}. A detailed description of the sample growth
and theoretical calculation can be found in the Supplemental Materials \cite{Supplemental}.

\begin{figure}
  \begin{center}
    \includegraphics[width=1\columnwidth]{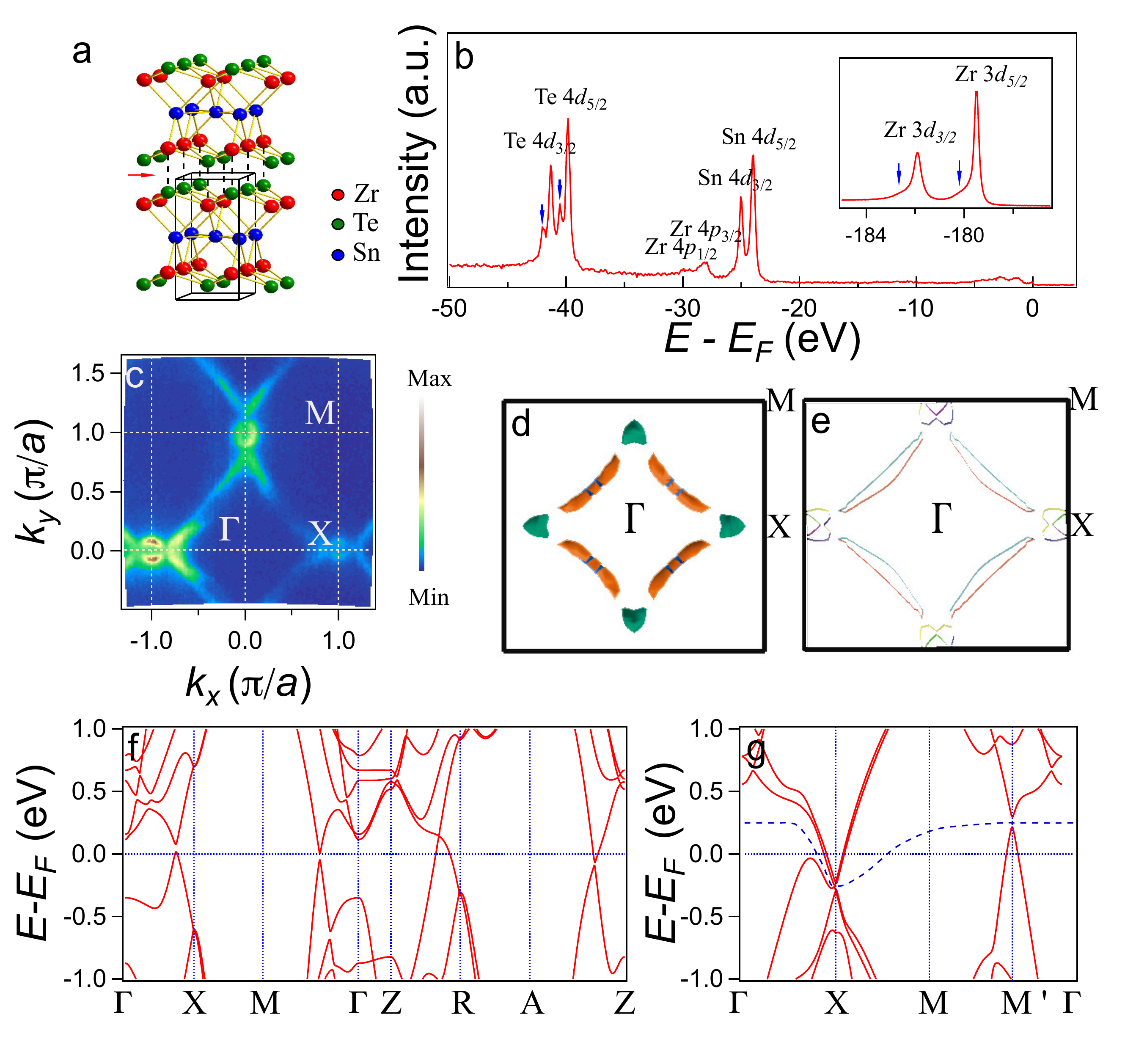}
  \end{center}
  \caption{(Color online) Crystal structure and electronic structure of {\ZST}.
    \textbf{a}, Crystal structure of {\ZST}. The arrow indicates that the cleavage takes place between the adjacent ZrTe layers, which
                breaks the weak Zr-Te bonds indicated as vertical dashed lines. The black cuboid demonstrates the structure of unit cell
                and also the monolayer used in calculation.
    \textbf{b}, Core level spectrum of {\ZST} recorded at photon energy $h\nu$ = 250 eV. The inset shows the magnification of the Zr
                3$d_{3/2}$ and 3$d_{5/2}$ peaks.
    \textbf{c}, FS intensity plot of {\ZST} recorded at $h\nu$ = 50 eV, obtained by integrating the spectral weight within $\pm$10 meV
                with respect to $E_F$.
    \textbf{d}, Calculated FSs of the 3D bulk crystal in top view along (001).
    \textbf{e}, FSs of a free-standing monolayer {\ZST} from first-principles calculations.
    \textbf{f,g}, Calculated band dispersions along the high-symmetry lines for bulk and monolayer {\ZST}, respectively. The blue dashed
                  curve in \textbf{g} represents the fictitious Fermi level.
  }  \label{F1}
\end{figure}

{\ZST} crystallizes in a PbFCl-type crystal structure with space group $P$4/$nmm$ \cite{Wang1995}, in which Sn is located at the
center of a tetrahedron consisting of Zr atoms, as illustrated in Fig. 1a. It is isostructural to the well known `111' type iron-based
superconductor LiFeAs \cite{Wang2008}. The relatively weak Zr-Te bonding between two neighboring slabs provides a natural cleavage plane
between the adjacent ZrTe layers, which yields a (001) surface with Zr and Te termination. This is consistent with the core level spectrum
in Fig. 1b, in which the double peaks of Te 4$d_{3/2}$ and 4$d_{5/2}$ split further into a total of four peaks. In the inset of Fig. 1b,
the Zr 3$d_{3/2}$ and 3$d_{5/2}$ peaks exhibit shoulders on the higher binding energy side. These indicate that the chemical environments
of the Te and Zr ions in the terminating layer are different from those inside. The measured Fermi surfaces (FSs) in Fig. 1c consist of
small electron pockets at $X$ and ``lenses''-like hole pockets in the $\Gamma$--$M$ direction. Such a FS topology is distinctly different
from the calculations of bulk {\ZST} in Fig. 1d, in which there is no FS centered around $X$. In contrast, it looks very similar
to that of monolayer {\ZST} in Fig. 1e.

The most distinct difference in the calculated electronic structures between the bulk (Fig. 1f) and the free-standing monolayer
(Fig. 1g) is that the electron bands at $X$ are shifted down by $\sim$1 eV, forming two electron pockets centered at $X$ in the
monolayer. This arises from the dangling bond of the Zr 4$d_{z^2}$ and 4$d_{xz}$+4$d_{yz}$ orbitals in the monolayer (see Supplemental
Materials \cite{Supplemental}). In bulk {\ZST}, these Zr 4$d$ orbitals are bonded with the Te 5$p_z$ and Zr 4$d$ orbitals in the
adjacent unit cell along the (001) direction. As discussed in Ref.~\onlinecite{Xu2015}, the Dirac cone-like bands around $E_F$ in
both the $k_z$ = 0 and $\pi$ planes will open gap when SOC is included since these bands have the same irreducible representation.
Similarly, SOC opens band gaps at $X$ ($\sim$75 meV) and $M^\prime$ ($\sim$80 meV) on the path $\Gamma$--$M$ in the monolayer {\ZST}
in Fig. 1g. As the gaps are located at different energies, we use a dashed curve to represent the fictitious Fermi level. Thus, the
$Z_2$ number is well defined for the bands below the curved Fermi level since there is a finite band gap at each $k$-point. It is
determined to be 1 by counting the parity of all occupied states at four time-reversal-invariant momenta. These indicate that the
monolayer {\ZST} could be a 2D TI once the curved Fermi level is straightened by band manipulation as discussed in the following.

\begin{figure}[b]
  \begin{center}
    \includegraphics[width=0.9\columnwidth]{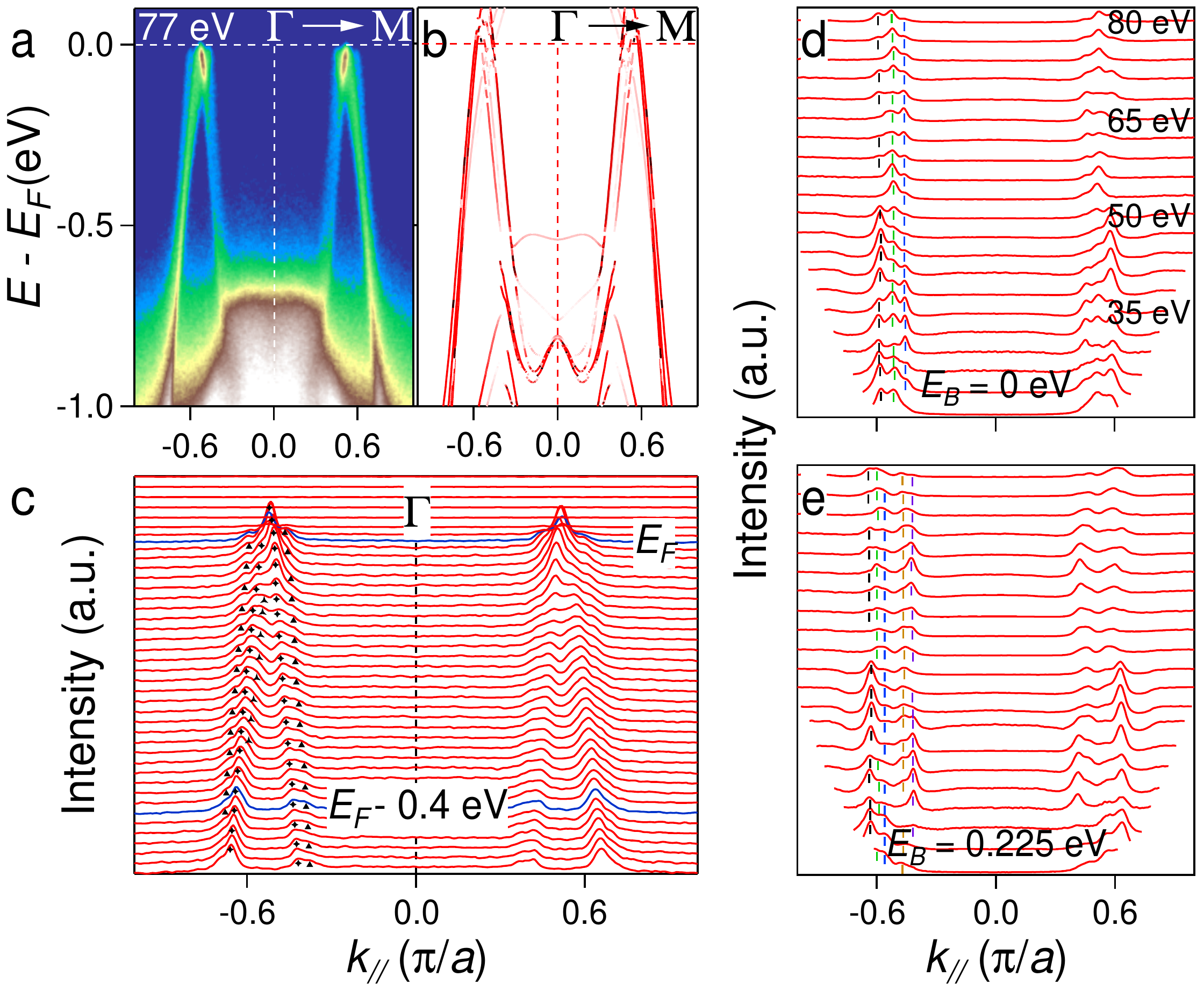}
  \end{center}
  \caption{(Color online) Band structure along $\Gamma$--$M$.
    \textbf{a}, Photoemission intensity plot along $M$--$\Gamma$--$M$ with $h\nu$ = 77 eV.
    \textbf{b}, Calculated band structure along $M$--$\Gamma$--$M$ for a seven-unit-cell thick slab. The intensity of the red color
                scales the spectral weight projected to the top two unit cells.
    \textbf{c}, MDC plot of \textbf{a}.
    \textbf{d,e}, MDC plots of the photon-energy dependent spectra at $E_F$ and $E_B$ = 225 meV, respectively. The dots and dashes
                  in \textbf{c}, \textbf{d}, and \textbf{e} are extracted peak positions, serving as guides to the eye.
  } \label{F2}
\end{figure}

In order to illuminate the topological character of the measured electronic structure, we have investigated the band dispersions along
the high-symmetry lines $\Gamma$--$X$, $X$--$M$, and $\Gamma$--$M$. The band structure along $\Gamma$--$M$ is shown in Fig. 2. The intensity
plot in Fig. 2a exhibits a Dirac-like band structure with crossing points near $E_F$. The cone-like feature is consistent with the monolayer
and bulk calculations. From the momentum distribution curves (MDCs) in Fig. 2c, one can see that each branch of the cone consists of two or
three nearly parallel bands. The multiple bands feature was reproducible in several measured samples, which rules out the possibility of
extrinsic effects such as multiple terraces on the cleavage surfaces or faults in the crystals. The feature is further confirmed by the
multiple peaks in the MDCs taken with different photon energies (Figs. 2d and 2e). The extra bands are not reproduced by either the monolayer
or bulk calculations. To understand the experimental observation, we have performed a slab model calculation for a thickness of seven unit
cells along the $c$ lattice. The slab model calculation is a good way to simulate the real sample situation in ARPES measurements \cite{
Lv2015,Lv2015a}. The calculated band dispersions with spectral weight from the top two unit cells are plotted in Fig. 2b, which can reproduce
the experimental bands very well and indicate that the ARPES experiments detect the signals of escaped photoelectrons mainly from the top two
unit cells. The outmost Zr atoms have different chemical environment from those inside, causing the slightly split parallel bands.

\begin{figure}
  \begin{center}
    \includegraphics[width=1\columnwidth]{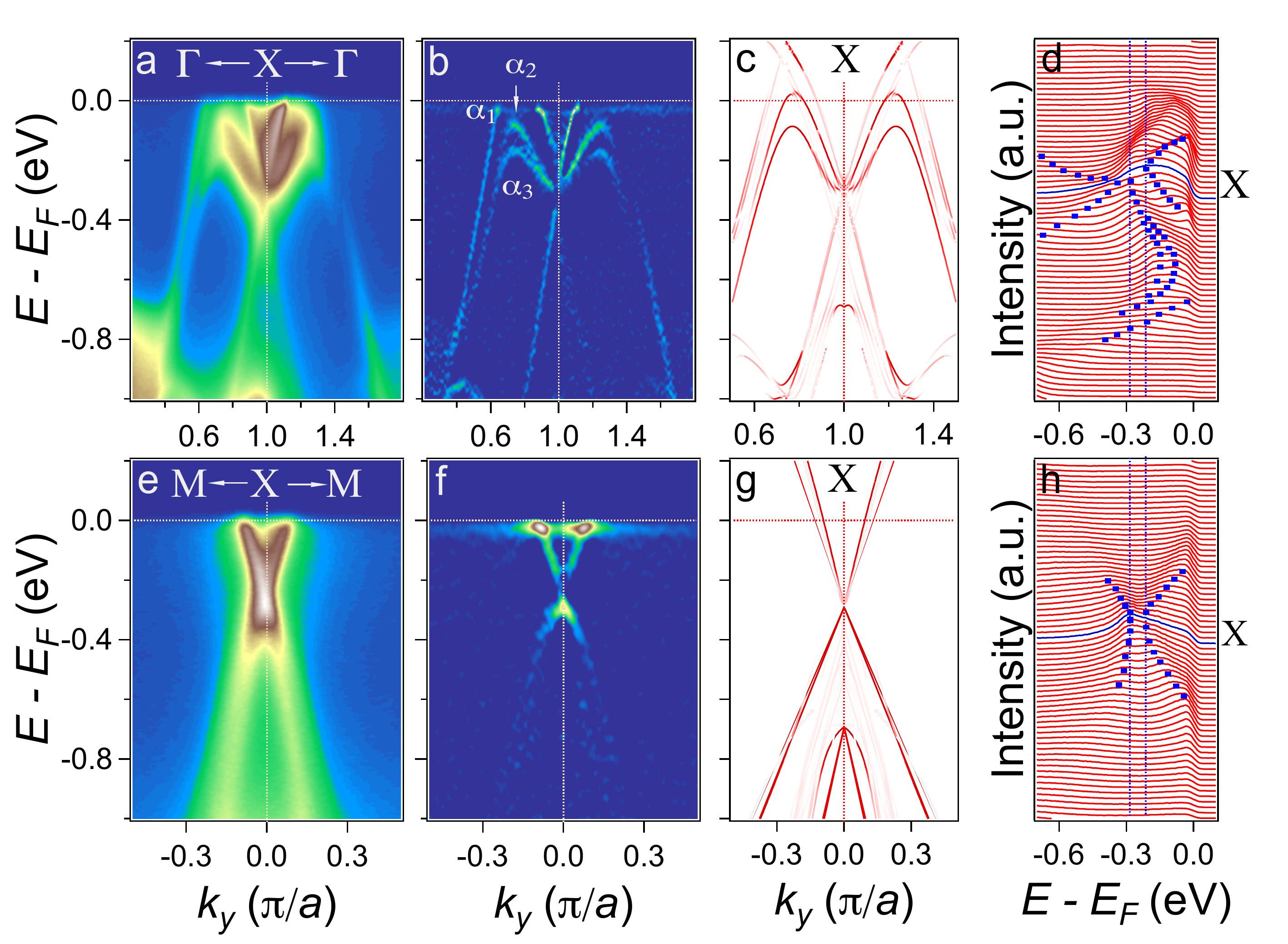}
  \end{center}
  \caption{(Color online) Band structure around $X$.
   \textbf{a}, Photoemission intensity plot along $\Gamma$--$X$--$\Gamma$ with $h\nu$ = 55 eV.
   \textbf{b}, 2D curvature intensity plot of \textbf{a}.
   \textbf{c}, Calculated band structure along $\Gamma$--$X$--$\Gamma$ for a seven-unit-cell thick slab. The intensity of the red
               color scales the spectral weight projected to the top two unit cells.
   \textbf{d}, Energy distribution curves (EDCs) of \textbf{a}. The blue dots are extracted peak positions, serving as guides to
               the eye. The vertical dashed lines indicate the gap at $X$.
   \textbf{e--h}, Same as \textbf{a--d} but along $M$--$X$--$M$.
  } \label{F3}
\end{figure}

\begin{figure}
  \begin{center}
    \includegraphics[width=1\columnwidth]{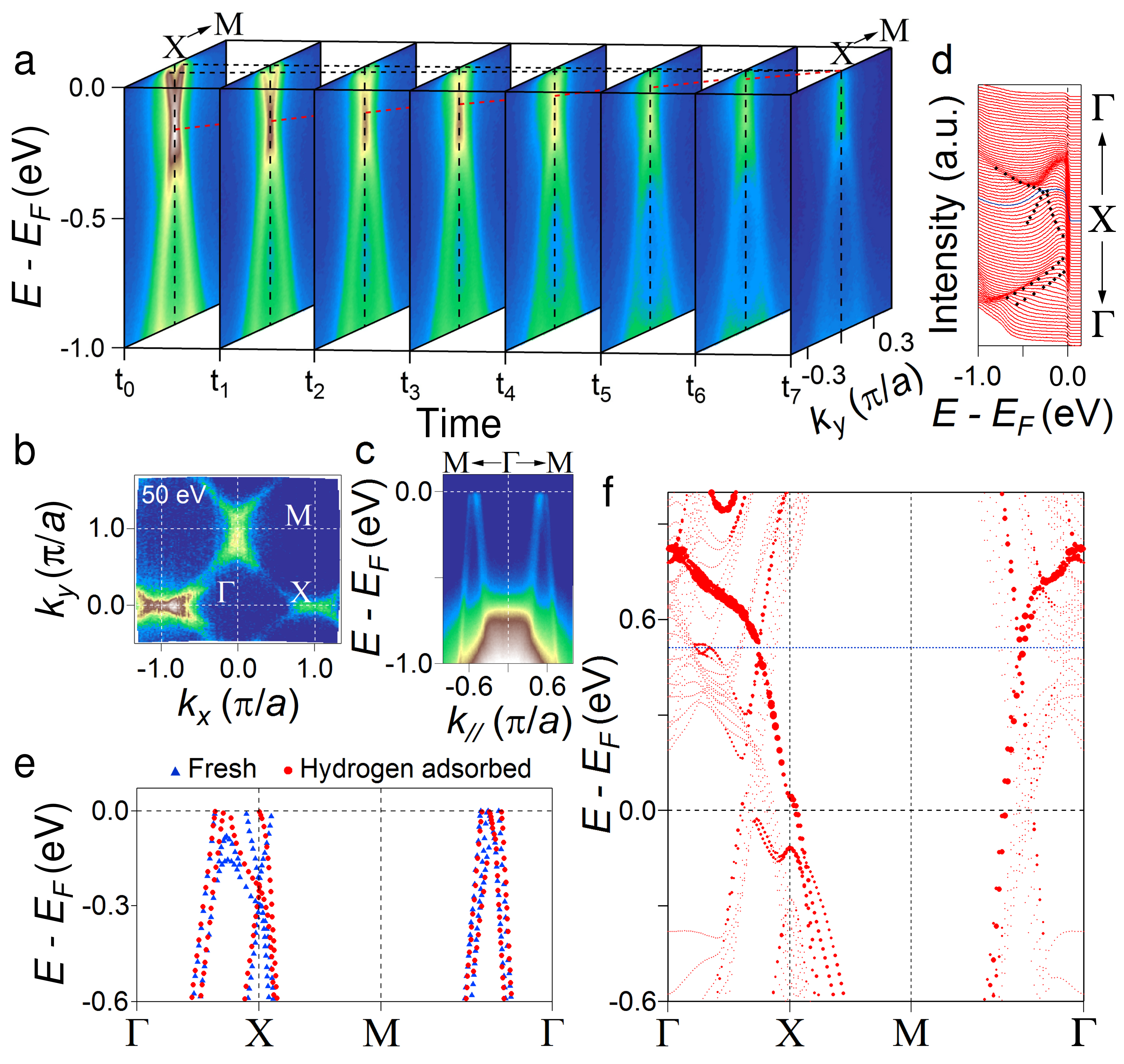}
  \end{center}
  \caption{(Color online) Manipulate the band structure with hydrogen adsorption.
   \textbf{a}, Evolution of the band dispersions along $M$--$X$--$M$ as a function of the exposure time in vacuum at $T$ = 20 K. $t_0$--$t_7$
               is the time sequence of the spectra, being separated with each other by $\sim$2 hours. $t_0$ and $t_7$ correspond to a fresh
               and a moderate hydrogenated surface, respectively. The top two black dashed lines indicate the crossing points of the electron
               band at $E_F$. The red dashed line is the trace of the band gap at $X$.
   \textbf{b}, FS intensity plot at $t$ = $t_7$ ($h\nu$ = 50 eV) by integrating the spectral weight within $\pm$10 meV with respect to
               $E_F$. The intensity at $X$ originates from the valence band top near $E_F$.
   \textbf{c}, Photoemission intensity plot along $M$--$\Gamma$--$M$ at $t$ = $t_7$.
   \textbf{d}, EDC plot along $\Gamma$--$X$--$\Gamma$ at $t$ = $t_7$. The band dispersions are marked by black dots.
   \textbf{e}, Extracted band dispersions along the high-symmetry lines of the ``fresh'' surface (blue solid triangles) and the surface
               with hydrogen adsorption at $t$ = $t_7$ (red solid circles).
   \textbf{f}, Calculated band structure along $\Gamma$--$X$--$M$--$\Gamma$ for a seven-unit-cell thick slab with all the surface Zr ions
               bonded with hydrogen atoms. The size of solid circles scales the spectral weight projected to the top one unit cell. The
               blue dashed line indicates the opening of a global band gap.
  }  \label{F4}
\end{figure}

The band structures along $\Gamma$--$X$ and $X$--$M$ are summarized in Fig. 3. We clearly observe a non-degenerate Dirac cone-like band
structure with a band gap of $\sim$90 meV at $X$, as seen in Figs. 3d and 3h. This observation is well consistent with the monolayer
calculations presented in Fig. 1g. In addition to the Dirac cone-like bands, we distinguish three near-$E_F$ bands along $\Gamma$--$X$,
as seen in Fig. 3b. The outmost band ($\alpha_1$) disperses almost linearly and crosses $E_F$ at $k_y$ $\sim$ 0.64 $\pi$/$a$ while the
other two bands ($\alpha_2$ and $\alpha_3$) turn back at binding energy $E_B$ $\sim$ 80 and 150 meV, respectively. They are overlapped
at $E_B$ = 0.3 eV at $X$. Likewise, these experimental bands are well reproduced by the slab model calculation with spectral weight from
the top two unit cells in Figs. 3c and 3g. We note that the electron band, as well as the $\alpha_2$ and $\alpha_3$ bands, are mainly
contributed by the spectral weight from the top one unit cell. By summarizing the band dispersions along the high-symmetry lines, it is
concluded that the 3D {\ZST} crystals host a 2D electronic state in the topmost unit cell, which is close to that of the free-standing
monolayer.

As in the case of the monolayer, the dangling bond from the Zr 4$d$ orbitals in the topmost unit cell causes the down-shift
of the electron bands at $X$. One can manipulate these bands by modifying the chemical environment of these Zr through surface
decoration/coverage, or interface of heterostructure. Hydrogenation has been found as an effective method to modify the electronic
properties of graphene and silicene \cite{graphene3,silicene1}. We have investigated the effects of hydrogenation on these bands
of the topmost unit cell using the residual hydrogen gas, which is the main component in the main chamber of our ARPES system with
a pressure of 3$\times$10$^{-11}$ Torr, as measured by the residual gas analyzer. We find that the bands near $X$ associated with
the Zr 4$d$ dangling orbitals are sensitive to the hydrogen adsorption on the surface. As seen in Fig. 4a, the bands composed of
the dangling orbitals around $X$ steadily moves up with time, and eventually the valence band top reaches the vicinity of $E_F$
when keeping the samples at $T$ = 20 K. In contrast, the band dispersions along $\Gamma$--$M$ only change slightly in Fig. 4c.
The changes in the band structure are further confirmed by the FS intensity plot in Fig. 4b, in which the electron pocket at $X$
in Fig. 1c disappears while the ``lenses"-like hole pockets along $\Gamma$--$M$ are little changed.

Since the sensitivity of the energy gaps to the hydrogen adsorption at $X$ and $M^\prime$ is different, the energy levels of the
two gaps might become the same by controlling the hydrogen adsorption process. We have performed similar slab calculation but with
the surface Zr ions all bonded with hydrogen atoms. The band structure with spectral weight on the topmost unit cell is shown in
Fig. 4f. Firstly, the weighted band structure can well reproduce the most characteristic features observed in the hydrogenated
sample shown in Fig. 4e, such as the valence band top near $E_F$ around $X$ and the nearly unchanged hole pocket along $\Gamma$--$M$.
Secondly, there are electrons transferred from the surface Zr to the hydrogen atoms to form occupied bonding states, causing the
up-shift of the bands associated with the Zr 4$d$ orbitals. Thirdly, the two gaps along $\Gamma$--$X$ and $\Gamma$--$M$ are located
at the same energy level of $\sim$0.5 eV above $E_F$, although the band gap is quite small. We would like to stress that as discussed
in Ref.~\onlinecite{Xu2015}, these bands have the same irreducible representation in bulk. The symmetry is further lowered on the (001)
surface, which leads to definite band gap opening. Therefore, an ideal 2D TI with a global band gap in the topmost layer is to be
realized by such modification of the surface termination.

In conclusion, our results have confirmed the existence of 2D topological electronic bands on the surface of {\ZST} crystal. We
have also demonstrated that the surface decoration, such as hydrogenation, is a promising way to manipulating these bands to open
a global band gap. This brings us a high chance to achieve an ideal 2D TI, through the similar surface decoration or interface of
heterostructures, among the predicted $WHM$ family \cite{Xu2015}. Actually, our previous calculations have predicted that ZrSiO
and LaSbTe would be an ideal 2D TI with a global gap located at $E_F$ \cite{Xu2015}. The tunable 2D TI states in covered or
decorated top surface, as well as the fact that it is isostructural to the iron-based superconductor LiFeAs \cite{Wang2008},
bring great promises in fabricating superconductor-TI heterostructures, in which various extraordinary quantum phenomena, $e.g.$,
topological superconductivity and Majorana modes, could be induced by the superconducting proximity effect \cite{Qi2011,Fu2008}.

This work was supported by the Ministry of Science and Technology of China (Nos 2011CBA00108, 2012CB921701, 2013CB921700, and
2015CB921300), the National Natural Science Foundation of China (Nos 11274381, 11274359, 11422428, 11474340, and 11574394), and
the Chinese Academy of Sciences (No. XDB07000000).

\noindent{\it Note added.---}During the review of this paper, there have been six preprints \cite{ZrSiS1,ZrSiS2,ZrSiS3,ZrSiS4,ZrSiS5,
ZrSiS6} reporting the transport and ARPES measurements on ZrSiS, which shares the similar crystal structure and electronic structure
as {\ZST}.

\end{document}